\documentclass[preprint,showpacs,preprintnumbers,pre]{revtex4-1}

\usepackage{mathtools, physics, amsmath,amssymb,amsfonts,amstext,graphics,graphicx,subfigure,dcolumn,bm}
\usepackage{setspace}
\usepackage[colorlinks]{hyperref}
\usepackage{color}

\usepackage{mathbbol}
\usepackage{dirtytalk}
\usepackage{natbib}

\begin{document}

\title{A BBGKY-like Hierarchy for Quantum Field Theories}

\author{Michael H. Updike}
\email{michaelupdike@utexas.edu}
\author{Joshua W. Burby}
\email{jburby@lanl.gov}
\affiliation{Los Alamos National Laboratory, Los Alamos, New Mexico 87545, USA }
\date{\today}

\begin{abstract}
We present a Hamiltonian method of constructing BBGKY-like hierarchies for quantum field theories. With suitable choices, our method creates a hierarchical system of evolution equations for the k-th order reduced density matrices. These equations can be closed at finite order using methods developed for the classical BBGKY hierarchy to give non-perturbative approximations for the full quantum equations of motion. Classical observables can then be numerically computed from these approximate equations, providing an analytically tractable method of modeling high-energy environments where quantum effects play a pronounced role.
\end{abstract}

\maketitle

\section{Introduction}
Current models of interacting charged particles rely on approximation methods derived from classical electrodynamics. These schemes enjoy great success in relatively weakly interacting environments; however, their predictive power is wanting at higher energies. Large radiative corrections occur when particles begin emitting photons of energy comparable to their mass \cite{blackburn}. This is not to mention more exotic corrections from effects such as pair creation \cite{paircreation}. With pulsed laser facilities currently achieving radiation intensities of over $10^{23} W/cm^2$ \cite{plasmachallenges}, and next-generation burning plasma experiments set to probe comparable energy scales,  more sophisticated models must be developed to achieve predictive capabilities. These models must be semi-classical and therefore derive from quantum electrodynamics.

A common starting part for many classical reduced models is the Bogoliubov–Born–Green–\\Kirkwood–Yvon (BBGKY) hierarchy. The BBGKY hierarchy is a way of organizing the equations of motion for the $N$ particle distribution function as a hierarchical system of equations for the $k \leq N$ point functions. Through various means, this hierarchy can be reduced to a small number of equations, providing workable approximations to the exact time evolution of a system. 

In \cite{MMW}, the BBGKY hierarchy was shown to have a Hamiltonian structure. In addition to being an important theoretical revelation, this structure offered insight into how BBGKY hierarchies could be constructed for other Hamiltonian theories. Indeed, in \cite{Nparticle}, and separately in \cite{otherpaper}, a BBGKY-like hierarchy was obtained for $N$ boson quantum mechanics. Using a filtration on quantum operators, as opposed to phase space functions, both papers obtained equations for the $k$th order reduced density matrices very similar to those of the classical BBGKY hierarchy.

In this paper, we generalize the constructions of \cite{MMW}, \cite{Nparticle}, and \cite{otherpaper} to arbitrary quantum field theories. Using a filtration of Hermitian polynomials in creation/annihilation, we are able to derive a Hamiltonian BBGKY-like hierarchy for the field-theoretic reduced-density matrices. Unlike both the classical and $N$-boson hierarchies, our hierarchy has infinitely many equations. Further, compared to the simple coupling of the other hierarchies, our hierarchy generically couples the $k$th order variable to the $k+1$ through $k+4$th order variables. Despite these differences, methods to truncate the classical BBGKY hierarchy can be applied to our hierarchy without issue, rendering our equations small in number and thus computable. 

To the best of our knowledge, the hierarchy we present is novel. We note, however, that a hierarchy of n-particle Wigner functions was constructed for $\lambda \phi^4$ theory in \cite{scalar}. While quite different from the approach we employ here, it remains open whether a connection between our hierarchies exists.

With our work, we hope to kickstart the field of using geometric mechanics and quantum field theory to study high-energy-density, many-body systems. Such a viewpoint seems not only natural but also powerful, as witnessed by the relative brevity of our constructions. With further investigation along these lines, we trust that a much greater understanding of high-energy-density environments can be achieved.

\section{Overview} 
In $\ref{sec:3}$, we generalize the works of \cite{MMW}, \cite{Nparticle}, and \cite{otherpaper} to provide a  framework from which a wide class of quantum field theoretic BBGKY-like hierarchies may be constructed. These hierarchies preserve the Hamiltonian structure of the field theory and take a form almost identical to the classical BBGKY hierarchy.
In \ref{sec:4}, we specialize our discussion to the hierarchy formed from the filtration of ladder operator polynomials, which we dub the canonical hierarchy. We show that the canonical hierarchy has many desirable properties, making it particularly well-suited to approximation schemes. In particular, relevant observables such as the energy and spectral densities depend only on the lowest-level hierarchy variables and are thus easily computed without recourse to further approximations. We demonstrate this in \ref{sec:5}. Finally, in \ref{sec:6}, we offer several avenues of further work including a method of obtaining Hamiltonian closures to our hierarchy.

\section{The Hamiltonian Theory of QFT Heirachies} \label{sec:3}
For an arbitrary quantum field theory, we let $\mathfrak{g}$ denote the space of hermitian operators. This space forms a lie-algebra under the usual commutator bracket $-i[\cdot,\cdot]$. A distinguished subset of the dual space $\mathfrak{g}^*$ is the space of trace-class, positive-semidefinite, hermitian operators $\mathfrak{d}$ dualized under the trace map $\rho \in \mathfrak{d}: A \mapsto \left<\rho,A\right> \coloneqq \text{Tr}(A\rho)$. By normalizing the trace, we may identify every nonzero element of $\mathfrak{d}$ with a density matrix. Provided we restrict to unitary equations of motion, $\mathfrak{d}$ can therefore be thought of as the phase space of the quantum theory.

We equip $\mathfrak{d}$ with a Poisson structure by defining the bracket $\{ \cdot, \cdot \} : C^\infty(\mathfrak{d}) \times C^\infty(\mathfrak{d}) \to C^\infty(\mathfrak{d})$
\begin{equation}
\{F,G\}(\rho) = \left<\rho, -i \left[\frac{\delta F}{\delta \rho},\frac{\delta G}{\delta \rho}\right] \right>,
\end{equation}
where $\frac{\delta F}{\delta \rho}$ is the unique element of  $\mathfrak{g}$ satifying 
\begin{equation}
\frac{d}{d \epsilon}|_{\epsilon = 0} F(\rho + \epsilon \delta \rho) = \left< \delta \rho, \frac{\delta F}{\delta \rho}\right>. 
\end{equation}
Given a Hamiltonian $H \in \mathfrak{g}$, the Hamiltonian flow of $H$ (viewed as a function on $\mathfrak{d}$) provides the usual quantum mechanical equations of motion.
Indeed, if $\rho(t): J \subset \mathbb{R} \to \mathfrak{d}$ integrates the Hamiltonian flow, then for any $F \in C^\infty(\mathfrak{d})$,
\begin{equation}
\frac{d}{d t} F(\rho(t)) = \left< \frac{\delta F}{\delta \rho}, \frac{d}{dt}\rho(t) \right> =  \{F, H\}(\rho(t)) = -i\left<\rho(t),\left[ \frac{\delta F}{\delta \rho}, H\right]\right> = i Tr\left(\frac{\delta F}{\delta \rho}\left[ \rho(t), H\right]\right),
\end{equation}
where the first equality follows from the chain rule and the last from the cyclic property of the trace. It is easy to see that this equality holds for all $F \in C^\infty(\mathfrak{d})$ if and only if 
\begin{equation}
\frac{d}{dt}\rho(t) = i[\rho(t), H],
\end{equation}
which is the desired relation.

Let $\mathfrak{g}_1 \subset \mathfrak{g}_2 \subset \mathfrak{g}_3 \subset \hdots  \subset \mathfrak{g}$ be a sequence of operator subspaces such that $-i[\mathfrak{g}_n, \mathfrak{g}_m] \subset \mathfrak{g}_{n+m-1}$ and $\bigcup_i \mathfrak{g}_i$ is dense in $\mathfrak{g}$. We call such a sequence a filtration. For any filtration, we define 
\begin{equation}
\mathfrak{G} \coloneqq \bigoplus_{i =1}^\infty \mathfrak{g}_i.
\end{equation}
We equip $\mathfrak{G}$ with the lie-bracket
\begin{equation}
[(A_i), (B_i)]_{\mathfrak{G}} = -i \left( [A_1, B_1] ,[A_2, B_1] + [A_1, B_2],\hdots, \sum_{n+m = k} [A_n, B_m], \hdots \right).
\end{equation}

As is the case for any direct sum, $\mathfrak{G}^* = \prod_{i = 1}^\infty \mathfrak{g}^*_i$.  Defining the subspaces \\$\mathfrak{d}_i = \{ \rho|_{\mathfrak{g}_i}: \rho \in \mathfrak{d}  \} \subset \mathfrak{g}_i^*$, we have the distinguished subspace of $\mathfrak{G}^*$ 
\begin{equation}
\mathfrak{D} \coloneqq \prod_{i = 1}^\infty \mathfrak{d}_i. 
\end{equation}
This space has the canonical Poisson bracket 
\begin{equation}
\{F,G \}_{\mathfrak{D}}((\rho_i)) = \left<(\rho_i), \left[\frac{\delta F}{\delta (\rho_i)}, \frac{\delta G}{\delta (\rho_i)}\right]_{\mathfrak{G}} \right>.
\end{equation}
Writing $\frac{\delta F}{\delta (\rho_i)} = (F_i) \in \mathfrak{G}$, this is equivalent to the formula 
\begin{align}
\left\{ F,G \right\}((\rho_i)) &= -i Tr\left(\rho_1[F_1, G_1]\right)  -iTr(\rho_2[F_2, G_1]) - iTr(\rho_2[F_1,G_2]) -iTr(\rho_3[F_3, G_1]) + \hdots. \\
& = i Tr\left(F_1 [\rho_1, G_1]\right)  +iTr(F_2[\rho_2, G_1]) + iTr(F_1[\rho_2,G_2]) + iTr(F_3[\rho_3, G_1]) + \hdots. \nonumber
\end{align}

Let $\alpha: \mathfrak{G} \to \mathfrak{g}$ denote the natural inclusion $\alpha: (A_i ) \mapsto \sum_{i} A_i $. One can easily verify that $\alpha$ is a lie-algebra homomorphism, and consequently that the dual function $\alpha^*: \mathfrak{d} \to \mathfrak{D},\, \alpha^*: \rho \mapsto (\rho|_{\mathfrak{g}_i})$ is a Poisson map. If $(H_i) \in \mathfrak{G}$ is any sequence of operators such that $\alpha((H_i)) = H$, and $\rho(t)$ evolves according to the flow on $\mathfrak{d}$, then $\alpha^*\rho(t)$ integrates the flow of $(H_i)$. This can be verified directly by checking that, for any $F \in C^\infty(\mathfrak{D})$,
\begin{equation}\label{eq:9}
    \frac{d}{dt} F(\alpha^*\rho(t)) = \left< (F_i), \alpha^*(\rho(t)) \right> = \left \{ F,(H_i) \right \}(\alpha^*\rho(t)).
\end{equation}

We may view the restricted dual operators $\rho|_{\mathfrak{g}_i}$ as equivalence classes $[\rho]_i$ of operators under the quotient $\rho_a \sim_i \rho_b$ if and only if $Tr(\rho_a A_i) = \rho(\rho_b A_i)$ for all $A_i \in \mathfrak{g}_i$. Letting $\rho_i \in [\rho]_i$ and $\pi_k: A \mapsto [A]_k$ be the quotient map, we define $[[\rho]_i,A_i] \coloneqq \pi_i[\rho_i, A_i]$ for any $A_i \in \mathfrak{g}_i$. Some simple algebra reveals that \eqref{eq:9} holds if and only if 
\begin{equation}\label{eq:10}
\frac{\partial}{\partial t}[\rho]_i = i[[\rho]_i, H_1] + i[[\rho]_{i+1}, H_2] + i[[\rho]_{i+2}, H_3] +\hdots.
\end{equation}
Provided that $H \in B_N$ for some small $N$, \eqref{eq:10} reveals that the equations for $[\rho]_i$ form a hierarchical structure. Furthermore, these equations are Hamiltonian. Truncating or otherwise closing the hierarchy, we obtain approximate equations for  $[\rho]_1, \hdots, [\rho]_N $. Approximate expectation values are then obtained for every operator $O$  such that $Tr(O \rho)$ depends only on the equivalence class $[\rho]_N$.
\section{The Canonical Hierarchy}\label{sec:4}
Given some set of operators $\{ X_\alpha \}$, the sequence of subspaces 
\begin{equation}
B_{k+1} = \{ \text{kth order Hermitian polynomials in } X_\alpha, X_\alpha^\dagger \}
\end{equation}
will always form a filtration of $\mathfrak{g}$. In particular, we will see it is desirable to set $\{ X_\alpha \} = \{a_{\mathbf{p},I} \}$, the annihilation operators of the quantum field theory. Here, $\mathbf{p}$ represents the particle momentum index, and $I$ represents any discrete indices (e.g. species, helicity, etc.). This filtration is particularly suited to approximations since, for any renormalizable theory, we may write that $H = \sum_{i = 2}^{5}H_i$ with $H_i \in B_i - B_{i-1}$. Defining the reduced density matrices
\begin{equation}
\Gamma^{(m,n)}_{(I'_1,\hdots,I'_n, I_1,\hdots,I_m)}[\rho](\mathbf{p}_1,...,\mathbf{p}_m; \mathbf{p}'_1,...,\mathbf{p}'_n) \coloneqq \text{Tr}(\rho a_{\mathbf{p}'_1, I_1'}^\dagger...a_{\mathbf{p}'_n, I_n'}^\dagger a_{\mathbf{p}_1, I_1}...a_{\mathbf{p}_m, I_m}), 
\end{equation}
this choice of filtration further allows the equivalence classes $[\rho]_{k}$  to be linearly encoded into a finite collection $\Gamma_k[\rho] = (\Gamma^{(m,n)}_{\mathcal{I}}[\rho]: m+n <k )$, say by the map $\epsilon_k: [\rho]_k \mapsto \Gamma_k[\rho]$. One easily verifies that $\epsilon_k$ is injective and hence invertible on its image. Since applying $\epsilon_k$ commutes with time derivatives, the equations of motion for $\Gamma_k$ can be obtained from $\eqref{eq:10}$ as 
 \begin{equation}\label{eq:13}
     \frac{\partial}{\partial t}\Gamma_i = i\epsilon_i  [\epsilon^{-1}_{i+1}(\Gamma_{i+1}), H_2] + i\epsilon_i [\epsilon^{-1}_{i+2}(\Gamma_{i+2}), H_3] + i \epsilon_i[\epsilon^{-1}_{i+3}(\Gamma_{i+3}), H_4] + i\epsilon_i [\epsilon^{-1}_{i+4}(\Gamma_{i+4}), H_5].
 \end{equation}
 
The power of these equations lies in the fact that they are hierarchical and can thus be closed at finite order using methods originally developed for the classical BBGKY hierarchy. The simplest such method is to set $\Gamma_i \equiv 0$ for $i >N$. In this case, the approximate equations of motion read 
\begin{align}
     &\frac{\partial}{\partial t}\Gamma_1 = i\epsilon_i  [\epsilon^{-1}_{i+1}(\Gamma_{2}), H_2] + i\epsilon_i [\epsilon^{-1}_{i+2}(\Gamma_{3}), H_3] + i \epsilon_i[\epsilon^{-1}_{4}(\Gamma_{i+3}), H_4] + i\epsilon_i [\epsilon^{-1}_{5}(\Gamma_{i+4}), H_5],
     \\
    \;\;\;\;\;\;\; \vdots \nonumber
   \\
    &\frac{\partial}{\partial t}\Gamma_{N-3} = i\epsilon_i  [\epsilon^{-1}_{N-2}(\Gamma_{2}), H_{N-2}] + i\epsilon_i [\epsilon^{-1}_{N-1}(\Gamma_{3}), H_{N-1}] + i \epsilon_i[\epsilon^{-1}_{4}(\Gamma_{N}), H_4],  \nonumber
    \\
    &\frac{\partial}{\partial t}\Gamma_{N-3} = i\epsilon_i  [\epsilon^{-1}_{N-2}(\Gamma_{2}), H_{2}] + i\epsilon_i [\epsilon^{-1}_{N-1}(\Gamma_{3}), H_{3}] + i \epsilon_i[\epsilon^{-1}_{4}(\Gamma_{N}), H_4],  \nonumber
\\
  &\frac{\partial}{\partial t}\Gamma_{N-2} = i\epsilon_i  [\epsilon^{-1}_{N-1}(\Gamma_{2}), H_{2}] + i\epsilon_i [\epsilon^{-1}_{N}(\Gamma_{3}), H_{3}],  \nonumber
\\
  &\frac{\partial}{\partial t}\Gamma_{N-1} = i\epsilon_i  [\epsilon^{-1}_{N}(\Gamma_{2}), H_{2}],  \nonumber
\\
   &\frac{\partial}{\partial t}\Gamma_{N} = 0. \nonumber
\end{align}
A more sophisticated method involves cluster expanding the reduced density matrices into their correlations below the $N$th order. This scheme similarly reduces $\eqref{eq:13}$ to $N$ equations. However, this method represents a superior weak-interaction limit since any relativistic quantum field theory has asymptotically vanishing correlations. In the simplest case of a scalar theory and $N =1$, this amounts to approximating 
\begin{equation}\label{above}
    \Gamma^{(m,n)}(\mathbf{p}_1,...,\mathbf{p}_m;\mathbf{p}'_1,...,\mathbf{p}'_n) \approx  \overline{\Gamma^{(1,0)}}(\mathbf{p}_1')\hdots\overline{\Gamma^{(1,0)}}(\mathbf{p}_n')\hdots \Gamma^{(1,0)}(\mathbf{p}_1) \hdots \Gamma^{(1,0)}(\mathbf{p}_m).
\end{equation}
Substituting this expression into $\eqref{eq:13}$ for $i=1$, we obtain an equation of the form $\partial_t \Gamma_1 = F(\Gamma_1)$ which can be solved and substituted into $\eqref{above}$. 

A major fault in both of these methods is that they do not produce Hamiltonian equations of motion, and hence break the underlying structure of the quantum theory. Ideally, approximations to  \eqref{eq:13} should close the hierarchy in a Hamiltonian manner. However, a general method for doing so remains open. In \ref{sec:6}, we offer several possible routes toward this goal. 

We note that once an approximation method has been chosen, the equations of motion are most effectively computed using the relation
\begin{equation}\label{eq:16}
i\epsilon_i  [\epsilon^{-1}_{j}(\Gamma_{\mathcal{I}}^{(m,n)}), A] = -i\text{Tr}(\rho \, [a_{\mathbf{p}'_1, I_1'}^\dagger...a_{\mathbf{p}'_n, I_n'}^\dagger a_{\mathbf{p}_1, I_1}...a_{\mathbf{p}_m, I_m},A]).
\end{equation}
For example, consider a free scalar theory with the Hamiltonian  
\begin{equation}\label{eq:17}
H = \int \frac{d^3\mathbf{q}}{(2\pi)^3 (2E_{\mathbf{q}})} E_{\mathbf{q}} a^\dagger_{\mathbf{q}} a_{\mathbf{q}} \in B_3,
\end{equation}
where $E_{\mathbf{p}} = \sqrt{\mathbf{p}^2 + m^2}$ is the particle energy. We work in natural units and assume the commutation relations 
\begin{equation}
[a_{\mathbf{p},\mathcal{I}}, a^\dagger_{\mathbf{p'}, \mathcal{I}'}] = (2\pi)^3(2E_{\mathbf{p}}) \, \delta_{I,I'}\, \delta^{(3)}(\mathbf{p} - \mathbf{p}').
\end{equation}
A simple calculation reveals that 
\begin{equation}
\frac{\partial}{\partial t} \Gamma^{(n,m)}(t; \mathbf{p}_1,\hdots,\mathbf{p}_n; \overline{\mathbf{p}}_{1}\hdots, \overline{\mathbf{p}}_m) = 
\left( -i\sum_{j=1}^n \mathbf{p}_{j}t + i \sum_{j=1}^m \overline{\mathbf{p}}_j t \right) \Gamma^{(n,m)},
\end{equation}
with similarly trivial behavior for other free theories. 

For interacting theories,
$\eqref{eq:13}$ is no longer exactly solvable. If the scalar Hamiltonian was instead
\begin{equation}
H =  \frac{1}{2} \int \frac{d^3 \mathbf{q}}{(2\pi)^3(2E_\mathbf{q})} \frac{d^3 \mathbf{s}}{(2\pi)^2(2E_\mathbf{s})} h(\mathbf{q},\mathbf{s}) a_\mathbf{q}^\dagger a_\mathbf{s}^\dagger a_\mathbf{q} a_\mathbf{s} 
\end{equation}
with $h(\mathbf{q},\mathbf{s})$ a symmetric, real-valued function, then a similar computation reveals that 
\begin{align}
\frac{\partial}{\partial t} \Gamma^{(1,0)}(t;\mathbf{p}) &= - i\int \frac{d^3\mathbf{q}}{(2\pi)^3(2E_{\mathbf{q}})} \; \;  \; \Gamma^{(2,1)}(t;\mathbf{p},\mathbf{q}; \mathbf{q})h(\mathbf{p},\mathbf{q}), \\
\frac{\partial}{\partial t} \Gamma^{(1,1)}(t;\mathbf{p};\mathbf{p}') &= - i\int \frac{d^3\mathbf{q}}{(2\pi)^3(2E_{\mathbf{q}})} \; \; \Gamma^{(2,2)}(t;\mathbf{p},\mathbf{q}; \mathbf{p}',\mathbf{q})\left(h(\mathbf{p},\mathbf{q}) - h(\mathbf{p}',\mathbf{q})\right), \nonumber \\
\vdots \nonumber
\end{align}
which is no longer trivial, highlighting the general need to seek approximations to $\eqref{eq:13}$. 
\section{Computing Observables}\label{sec:5}
 Perhaps the simplest observable is the expected density of particles with discrete indices $I$ and momentum $\mathbf{p}$,
\begin{equation}
    D_I(\mathbf{p}) \coloneqq \left< \frac{1}{(2\pi)^3 2E_{\mathbf{p}} }a_{\mathbf{p},I} ^\dagger a_{\mathbf{p},I}\right> =\frac{1}{(2\pi)^3 2E_{\mathbf{p}}} \Gamma_{(I,I)}^{(1,1)}(\mathbf{p}; \mathbf{p}).
\end{equation}

Another observable of interest is the total energy stored in a region of space. For any relativistic theory, we may write the Hamiltonian as
\begin{equation}
H = \int d^3\mathbf{x} \; \mathcal{H} (\mathbf{x}).
\end{equation}
where $\mathcal{H}(\mathbf{x})$ is the $(0,0)$ component of the physical stress-energy tensor. The energy distribution of a system is then
\[
\mathcal{E}(\mathbf{x}) \coloneqq \left<\mathcal{H}(\mathbf{x}) \right> = F(\Gamma_5).
\]
where $F$ is some function depending on the Hamiltonian.
For interacting theories, $\left<\mathcal{H}(\mathbf{x}) \right>$ is in general a very complicated function. In many cases, however, the energy stored in particle interactions is much smaller than the energy of the particles themselves. For such cases, we may approximate 
\begin{equation}
    \left< \mathcal{H}(\mathbf{x}) \right> \approx \left< \mathcal{H}_{\text{free}}(\mathbf{x})\right> = F_{\text{free}}(\Gamma_3),
\end{equation}
with $\mathcal{H}_{\text{free}}$ being the Hamiltonian density of the free theory.
For example, for an arbitrary scalar theory, we may use $\eqref{eq:17}$ to approximately compute that
\begin{align}\label{eq:}
 \mathcal{E}(\mathbf{x}) & \approx \int \frac{d^3\mathbf{k}}{(2\pi)^3(2E_\mathbf{k})} \;  \frac{d^3\mathbf{p}}{(2\pi)^3(2E_\mathbf{p})} \;\\ & (m^2- E_{\mathbf{p}} E_\mathbf{k}+ \mathbf{p} \cdot \mathbf{k} ) e^{i\mathbf{x}\cdot (\mathbf{k}+ \mathbf{p})}\Gamma^{(2,0)}(t; \mathbf{k},\mathbf{p})\nonumber + e^{-i\mathbf{x} \cdot (\mathbf{k} - \mathbf{p})} (m^2+ E_{\mathbf{p}} E_\mathbf{k}+ \mathbf{p} \cdot \mathbf{k} ) \Gamma^{(1,1)}(t;\mathbf{p};\mathbf{k}). \nonumber
\end{align}
 For a massive theory, we may further consider the limit whereby the particle mass is much larger than the particle momenta, and hence that the energy of the fields resides entirely in the mass energy of the particles. Letting $\left< A \right>_{\mathbf{p} \to 0}$ denote the expectation value of an operator $A$ in such a limit, the number density of particles can thus be computed as
\begin{equation}
N(\mathbf{x}) = \frac{\left< \mathcal{H}_{\text{free}}(\mathbf{x}) \right>_{\mathbf{p} \to 0}}{m}.
\end{equation}
In the case of a scalar theory, for example, the expected particle density is
     \begin{align}\label{eq:26}
 N(\mathbf{x})  =  \int \frac{d^3\mathbf{k}}{(2\pi)^3 \sqrt{2E_\mathbf{k}}}  \frac{d^3\mathbf{p}}{(2\pi)^3 \sqrt{2E_\mathbf{p}}}  e^{-i\mathbf{x} \cdot (\mathbf{k} - \mathbf{p})} \Gamma^{(1,1)}(t;\mathbf{p};\mathbf{k}),
\end{align}
which is verified as the expectation value of the operator
\begin{equation}\label{eq:27}
\mathcal{N}(\mathbf{x}) =  \int \frac{d^3\mathbf{k}}{(2\pi)^3 \sqrt{2E_\mathbf{k}}}  \frac{d^3\mathbf{p}}{(2\pi)^3 \sqrt{2E_\mathbf{p}}} e^{-i\mathbf{x}\cdot(\mathbf{k} - \mathbf{p})} a_\mathbf{k}^\dagger a_{\mathbf{p}}. 
\end{equation}
Indeed, the interpretation of $\mathcal{N}(\mathbf{x})$ as a number-density is confirmed upon checking that $\int d^3\mathbf{x} \; \mathcal{N}(\mathbf{x})$ is the usual number operator. Similar operators for other theories are obtained from \eqref{eq:27} by adding the appropriate discrete indices and external leg factors. 
\section{Discussion}\label{sec:6}
While an important first step, several more steps must be taken before our hierarchy can be used in the semi-classical modeling of high-energy environments. For one, before any computational work is undertaken, $\eqref{eq:13}$ must be recast as a differential equation and the desired initial conditions rewritten as reduced density matrices. This is straightforward in the simplest cases but requires considerable effort for realistic theories such as quantum electrodynamics. 
A more subtle issue is how nontrivial boundary conditions may be worked into our theory.  It is clear this involves working in position space, but we leave the details to future work. 

A more theoretical line of future work lies in finding Hamiltonian closures to our hierarchy equations. While it is desirable that approximations to the quantum evolution equations preserve the Hamiltonian structure of theory, it is not clear how this can be done. One promising line of investigation is to look for Poisson maps from some smaller Poisson manifold $P$ to $\mathfrak{D}$. Solutions to the Hamiltonian flow on $P$ would then push forward to solutions on $\mathfrak{D}$, while presumably being easier to solve. Given an operator ideal $\mathcal{I}$, a natural choice for $P$ would be the subspace of $\mathfrak{D}$ consisting of sequences $(\rho_1, \rho_2,\hdots)$ such that $\rho_i(\mathfrak{g}_i \cap \mathcal{I}) = 0$, with the Poisson bracket inherited from the quotient algebra $\mathfrak{g}/\mathcal{I}$. However, such ideals remain elusive if they exist. 

In a similar vein, one could look for Poisson structures on $\prod_{i=1}^N \mathfrak{d}_i$ close to the Poisson structure on $\prod_{i=1}^\infty \mathfrak{d}_i$. Although no longer an exact solution, solving for the flow on the smaller space would still serve as a Hamiltonian approximation to the full equations of motion.

\section{Acknowledgment}
\noindent  Research presented in this article was supported by the Los Alamos National Laboratory
LDRD program under project number 20230497ECR.
\\\\
LA-UR Number 23-31725
\nocite{*}
\bibliographystyle{abbrv}
\bibliography{sample}
\end{document}